# Observation of a uniaxial strain-induced phase transition in the 2D topological semimetal IrTe$_2$

C. W. Nicholson[1,*], M. Rumo[1], A. Pulkkinen[1,2], G. Kremer[1], B. Salzmann[1], M.-L. Mottas[1], B. Hildebrand[1], T. Jaouen[1,3], T. K. Kim[4], S. Mukherjee[4], K. Y. Ma[5], M. Muntwiler[6], F. O. von Rohr[5], C. Cacho[4] and C. Monney[1,*]

[1]*Université de Fribourg and Fribourg Centre for Nanoscience, Chemin du Musée 3, CH-1700 Fribourg, Switzerland*

[2]*School of Engineering Science, LUT University, FI-53850, Lappeenranta, Finland*

[3]*Univ Rennes, CNRS, IPR (Institut de Physique de Rennes) - UMR 6251, F-35000 Rennes, France*

[4]*Diamond Light Source, Harwell Campus, Didcot, OX11 0DE, United Kingdom*

[5]*Department of Chemistry, University of Zurich, Winterthurerstrasse 190, CH-8057 Zurich, Switzerland*

[6]*Paul-Scherrer-Institute, Forschungsstrasse 111, CH-5232 Villigen PSI, Switzerland*

[*]Corresponding author: christopher.nicholson@unifr.ch (C.W.N.); claude.monney@unifr.ch (C.M.)

**Strain is ubiquitous in solid-state materials, but despite its fundamental importance and technological relevance, leveraging externally applied strain to gain control over material properties is still in its infancy. In particular, strain control over the diverse phase transitions and topological states in two-dimensional (2D) transition metal dichalcogenides (TMDs) remains an open challenge. Here, we exploit uniaxial strain to stabilize the long-debated structural ground state of the 2D topological semimetal IrTe$_2$, which is hidden in unstrained samples. Combined angle-resolved photoemission spectroscopy (ARPES) and scanning tunneling microscopy (STM) data reveal the strain-stabilized phase has a 6x1 periodicity and undergoes a Lifshitz transition, granting unprecedented spectroscopic access to previously inaccessible type-II topological Dirac states that dominate the modified inter-layer hopping. Supported by density functional theory (DFT) calculations, we show that strain induces a charge transfer strongly weakening the inter-layer Te bonds and thus reshaping the energetic landscape of the system in favor of the 6x1 phase. Our results highlight the potential to exploit strain-engineered properties in layered materials, particularly in the context of tuning inter-layer behavior.**

Using external stimuli to manipulate the diverse phenomena observed in quantum materials may allow for tunable control over technologically relevant material properties. Within this context uniaxial strain has recently emerged as a powerful approach to influence the properties of solids[1–6] and offers a path to tailor both physical properties and device functionalities, particularly in the 2D TMDs[7–10]. While efforts to control phase transition behavior with strain have focused predominantly on oxide materials, there also exist many opportunities within the 2D semimetals, which routinely host multiple nearly degenerate structural, electronic and topological phases[11], thereby making them sensitive to external perturbation. In this regard, the family of layered tellurides are particularly promising[12], a prime example of which is 1$T$-IrTe$_2$. This high-atomic number material is predicted as a type-II bulk Dirac semimetal with a Dirac point slightly above the Fermi level[13] and presents first-order bulk phase transitions to a 5×1×5 structure at 280 K, and to an 8×1×8 structure at 180 K[14,15]. At the surface a complex staircase of nearly degenerate low-temperature phases with periodicity 3$n$+2 (i.e. 8×1, 11×1, 17×1…) coexist over scales of a few tens of nanometers[15,16]. All of the broken-symmetry phases display characteristic quasi-1D modulations typically identified as Ir dimers[15,17], although the changes to the in-plane bonding suggest a multi-center bond as a more complete description[18] (for brevity we will continue to use "dimers" throughout the text). The proposed ground state, a 6×1 phase[15,19–21], is typically observed only within nanoscale regions, making it all but inaccessible to most techniques. As a result, the electronic structure of the ground state, as well as any influence of the phase transitions on the bulk Dirac states, remains unclear, hindering efforts to elucidate the transition mechanism or exploit the topological properties. Substantial changes to the material behavior are produced by doping: superconductivity is induced by partial substitution of Ir with Pt[22] or Pd[23], or by temperature quenching[24], while partial substitution of Te with Se induces charge order[19,20,25], further emphasizing the metastable nature of the material. The range of competing phases observed in IrTe$_2$ strongly implies that its macroscopic behavior may be tunable via strain, allowing individual phases to be selectively stabilized without the need for external doping.

In this work, by applying a modest uniaxial tensile strain ($\varepsilon \sim 0.1\%$) to IrTe$_2$ single-crystals, we demonstrate the selective stabilization of a single structural phase transition with domain sizes four orders of magnitude larger than in unstrained samples. Complementary real and momentum space probes reveal this as a 6×1 charge ordered phase, a configuration that maximizes both the formation of Ir dimers and of Ir to Te charge transfer. We show that strain initiates this charge transfer already at room temperature, thereby removing the near degeneracy of the 3$n$+2 ladder of phases and favoring the 6×1 phase at low temperatures[18]. This energetic bias allows unprecedented spectroscopic access to the ground state of IrTe$_2$, including the previously unobserved bulk Dirac-like states, which undergo a Lifshitz transition due to the charge transfer. Concurrently, charge transfer results in a significant weakening of the majority of inter-layer Te bonds in the unit cell, resulting in a tenfold reduction of inter-layer hopping in the relevant states, and leaving the bulk Dirac states as the dominant inter-layer transport channel. These results demonstrate the power of strain to influence phase transitions, bonding and topology in the layered tellurides, and more broadly in the 2D semimetals.

Fig. 1a shows the hexagonal crystal structure, typical of the layered TMDs, in the high-temperature (HT) 1×1 phase of 1$T$-IrTe$_2$. In comparison, all low-temperature (LT) phases in IrTe$_2$, including the 6×1 (Fig. 1b), are characterized by the formation of Ir dimers stabilized by electronic energy gain[18,19]. The density of these dimers increases in each of the successive charge ordered phases, reaching a maximum in the 6×1 phase[19,26] which is generally considered as the ground state of the system. The band dispersion along the high-symmetry *LAL* direction of the bulk Brillouin zone (Fig. 1c), as obtained by angle-resolved photoemission spectroscopy (ARPES), is displayed in Fig. 1e for the HT phase in an unstrained sample. The corresponding Fermi surface (Fig. 1h) shows three-fold rotational symmetry consistent with the literature[26–29]. Upon cooling the unstrained sample (Fig. 1f), subtle changes occur due to the phase transitions in IrTe$_2$. The overall form of the electronic structure in the LT phase strongly resembles the HT phase, but with increased broadening of the bands and a reduction of spectral weight. The lack of clear features results from the presence of multiple domains and phases (see Fig. 2). While the disappearance of the three-fold symmetry in the Fermi surface (Fig. 1i) implies a dominant domain orientation within the probed region (50 μm), the absence of a single-phase domain hinders analysis of the resulting electronic structure. In dramatic contrast, the strained sample (Fig. 1g) displays a rich spectrum of remarkably sharp bands over a wide energy range, implying a uniform signal over the probed region originating from a single phase. Strain was applied along the *a*-axis of the HT phase using the home-built device pictured in Fig. 1d (see Methods). Of particular interest are sharp hole-like surface states and an apparent bulk-like hyperbolic dispersion close to $E_F$ (red arrows, Fig. 1g), discussed in more detail below, which are undiscernible in unstrained samples. The Fermi surface (Fig. 1j), reveals a clear directionality, breaking the rotational symmetry of the HT phase and resulting in a mirror-plane along the $k_x = 0$ line. Cuts along the $k_y$ direction (Supplementary Fig. S2) showing repeated surface and bulk states reveal this phase has a 6×1 in-plane periodicity, which is difficult to discern in Fig. 1j due to the small size of the repeated features and the variation of spectral weight.

The 6×1 periodicity of the strain-induced state is confirmed in Fig. 2, which demonstrates the effect of strain on the real-space surface structure as revealed by low-temperature scanning tunneling microscopy (STM) measurements. Unstrained samples display a mixture of differently oriented domains (Fig. 2a) which form due to the three-fold degeneracy of the HT phase. Within these rotational domains, there exist multiple phases with different 3n+2 periodicities (Fig. 2d). Again, in contrast, strained samples reveal a clear uni-directional domain (Fig. 2b), with a single 6×1 periodicity (Fig. 2f). The line cut in Fig. 2g shows that two non-identical groups of three atoms comprise the 6×1 periodicity in agreement with previous work[15,20]. Although the individual STM images are limited in size, the same 6×1 phase – always with the same orientation – is found in multiple images across the sample surface over hundreds of microns: more than half of the sample area (Supplementary Fig. S3). This macroscopic domain size is also seen in low energy electron diffraction (LEED), which averages over a region of similar dimensions (Fig. 2c and e). In the strained case, we indeed record a single domain orientation with 6×1 periodicity, contrasting the clear three-fold directionality of the unstrained sample. Further corroboration is obtained from micro-ARPES mapping (Fig. 2h), obtained by integrating the intensity of the sharp surface states characteristic of the 6×1 phase (Supplementary Fig. S4) across the sample surface. This reveals that the 6×1 phase is found over a continuous region of dimensions ~0.5x0.4 mm$^2$.

As is typical of phase transitions in the metal-chalcogenides[30] evidence of charge transfer is observed during the formation of the 6×1 phase. Due to the presence of polymeric bonds, the structure of IrTe$_2$ lies between that of pure 2D or 3D materials[12]. This results in an Ir$^{3+}$ configuration and hence a partial charge of Te$^{1.5-}$ on average in the HT phase. In the LT phase, a charge δ is transferred which produces modified Ir$^{3+\delta+}$ and Te$^{1.5-\delta/2-}$ species[19,26]. The electronic energy gain from Ir dimer formation[17] competes with the lattice deformation energy[31], making a complete dimerization of the surface energetically unstable[19]. As a result, both $Ir^{3+}$ and $Ir^{3+\delta+}$ charge species are present in the LT phase. This can be readily observed in X-ray photoemission spectroscopy (XPS) where the two distinct peaks appear in the Ir 4$f$ spectra[26,28] as shown in Fig. 3b. We note that the higher binding energy of the second peak implies reduced screening of the core potential, consistent with a reduced electronic density on the Ir atom ($Ir^{3+\delta+}$). Further evidence for a charge transfer is discussed below (Fig. 4). The ratio of the peak areas tracks the relative dimer density and indicates a mixture of phases in unstrained samples[26,29]. In contrast, for strained samples an increase of the $Ir^{3+\delta+}$ peak produces a ratio that accords perfectly with the expectation for a single 6×1 phase (0.67). Crucially, XPS measurements in the HT phase (Fig. 3a) reveal that a small population of the $Ir^{3+\delta+}$ species is already evident above the transition temperature in strained samples, which is absent for unstrained samples. This implies strain actually induces a charge transfer from Ir to Te, and that this is central to understanding the phase stabilization. The analysis of the peak ratio in the HT phase gives only 0.14, well below the ratio of 0.4 obtained in the 5x1 phase, which has the lowest dimer density of the ordered phases. An open question is whether the appearance of the second charge peak in the HT phase implies the existence of dimers above the phase transition temperature, and indeed whether an ordered phase can be induced at room temperature by increasing the strain level. We note, however, that dimer formation depends on the competition between electronic and lattice energy, and it is therefore possible in the strained system that a charge transfer is induced without dimer formation. Finally, we remark that no evidence for a continuous phase transition is observed in temperature dependent ARPES at this strain level (Supplementary Fig. S5).

In order to gain more insight into this redistribution of charge in the strain-stabilized phase, in Fig. 3c and d we compare the calculated charge distributions in the 1×1 and 6×1 phases, respectively. Particularly notable is that there is a clear increase of charge density in the inter-layer Te-Te region as a result of the phase transition, as charge is moved away from the $Ir^{3+}$ sites and onto the Te atoms. This implies the strain-induced charge transfer that produces the $Ir^{3+\delta+}$ signal in the HT phase also redistributes charge into the inter-layer region.

The impact of this charge transfer on the out-of-plane Te-Te bonds[25,32] is significant (the effect on the in-plane bonds has been addressed previously[18]). A particularity of the tellurides in comparison to other TMDs is the presence of 3D polymeric bonding structures[12], in place of the usual van der Waals gap. IrTe$_2$ indeed contains a network of weak interlayer covalent bonds in the HT phase[32]. In Fig. 4a we present a calculation in the 6×1 phase highlighting the different inter-layer bond lengths and their strengths relative to the HT phase. Bond strengths are obtained using the integrated crystal orbital Hamiltonian potential (ICOHP) method[33]. We find four inequivalent inter-layer Te-Te bonds in the 6×1 phase, compared with only one type in the HT phase (see Supplementary Table S1). Various bond strengths are changed in the 6x1 phase including a number which strengthen, in line with a multicenter bond description[18]. Three out of the four interlayer bonds are observed to weaken significantly across the phase transition, represented in blue in the figure (see also Supplementary Table S1). The reason for this is that the charge transferred from Ir$^{3+}$ populates anti-bonding Te-Te states in the inter-layer region, thereby reducing the overall bond strength. The resulting inter-layer bond weakening has been termed "depolymerization"[12,25,26], although quantities relevant for bonding such as the out-of-plane bond strengths and hopping have not previously been addressed in detail for the low-temperature phase. We provide a direct experimental quantification of the effect that the bond weakening has on the electronic structure and electronic hopping in the out-of-plane direction. We do so by comparing the out-of-plane ($k_z$) dispersion (Fig. 4b) in the HT and 6×1 phases, and reemphasize that it is only via the strain stabilization that we are able to access the electronic structure of the pure 6×1 phase. Between the LT and HT phases, the majority of states maintain their small out-of-plane dispersion. However, a sizeable change in the warping of the Fermi contour is observed for bulk states on either side of the Brillouin zone boundary, highlighted by the blue lines in the two panels of Fig. 4b. In general, such warped Fermi surface contours are characteristic of a strong anisotropy in the electronic hopping parameters and are routinely observed in low-dimensional materials. A small warping corresponds to a low coupling between chains (1D) or planes (2D) along the relevant real-space direction[34,35]. In the case of the present out-of-plane dispersion, the narrowing of the warping in the $k_z$ direction corresponds to a reduction of the inter-layer hopping in the LT phase. In contrast, significant dispersion is observed for these same states in the ($k_x$, $k_y$) plane, highlighting the quasi-2D behavior of IrTe$_2$ in the 6x1 phase. By applying a tight-binding model[34] (see Methods) we find the inter-layer hopping parameter to reduce by a factor of ten to only $t_c = -0.014$ eV in the LT phase (in comparison, the in-plane $t_a = -0.53$ eV). This prominent reduction of inter-layer coupling in the 6×1 phase therefore strongly enhances the 2D nature of the system. The observed layer decoupling further supports recent calculations that show monolayer IrTe$_2$ has an increased tendency towards the 6×1 phase[18] suggesting that the dimerized phases could potentially be stabilized at much higher temperatures by reducing the sample thickness to the monolayer limit.

These observations strongly implicate the inter-layer bond weakening in the mechanism of the strain-stabilized phase transition[25,26,32]. By inducing charge transfer, and hence inter-layer bond weakening already in the HT phase, strain reduces the amount of electronic energy that can be gained through dimerization of the Ir$^{3+\delta+}$ ions. This therefore destabilizes the nearly degenerate LT phases, and pushes the system to favor the formation of the phase with the highest dimer density and hence highest possible gain in electronic energy i.e. the 6×1 phase. In this way, the less stable $3n+2$ phases are removed from the LT phase diagram. The effect of strain is thus twofold: first, by defining a preferential direction, it breaks the degeneracy of the threefold dimer orientation[36]; second, it biases the energetic landscape of the system in favor of the 6×1 phase. Although the effects of strain are subtle in the HT phase, as expected for the perturbative strain level applied, they pave the way for the stabilized phase transition, with dramatic results at low temperatures. We note that similar bonding behavior is realized in a number of di- and tri-telluride materials spanning the 2D and 3D regimes[12], suggesting strain or electrical gating as powerful methods to control structural behavior and dimensionality in this class of materials.

In contrast to the discussion above, a state dispersing in $k_z$ appears around $k_x = 0$ in the LT phase (red, arrow, Fig. 4b). Due to its strong out-of-plane dispersion (Supplementary Fig. S6), this state is only observed around $k_z = 5$ Å$^{-1}$, i.e. the bulk A-point. This feature corresponds to the triangular block of states observed in Fig. 1g, and shown again at different $k_y$ positions in Fig. 4d. The strong $k_z$ dependence and "filled-in" nature of these states resulting from the projection of the bulk manifold reveals them as bulk states. Of note is that these bulk states have the cone-like hyperbolic dispersion of massless Dirac fermions (Supplementary Fig. S7), which occur ubiquitously in the group-10 TMDs and are predicted in IrTe$_2$[13]. However, the location of the type-II Dirac point at room temperature is above $E_F$, hence inaccessible to ARPES, while at low temperature the mixture of phases typically hides their true nature. The observed shift of the Dirac point to 350 meV below $E_F$ in the 6×1 phase occurs as the Dirac states are derived from the out-of-plane Te *5p$_z$* orbitals and hence are strongly doped by the charge transfer into Te states as described above. As we have shown, it is only possible to access these states spectroscopically in strained samples. By moving the Dirac point and the electron-like portion of the Dirac cone into the occupied states, strain produces a Lifshitz transition similar to the temperature-driven transitions observed in WTe$_2$[37] and ZrTe$_5$[38]. Such a dramatic change in Fermi surface topology is likely to have a significant impact on the transport properties in this material. In particular, the topological nature of the states involved in the transition may explain the observed large, non-saturating magnetoresistance[39] similar to the behavior in other layered di-tellurides[40–42]. The strain-stabilized order may even enhance such effects, paving the way to tunable magneto-resistive behavior. Given the out-of-plane character of the *5p$_z$* orbitals involved in the Dirac states, it is plausible that particularly large changes in inter-layer transport may be observed, and investigations of the resistivity anisotropy using e.g. focused ion beam methods[43] are highly desirable in this regard. The potentially topological nature of this out-of-plane transport makes IrTe$_2$ layers especially interesting for tuning inter-layer behavior in heterostructure architectures[44,45].

While a detailed discussion of the topological properties is beyond the scope of the current article, we nonetheless highlight a surprising observation regarding these Dirac-like states: the dispersion is not compatible with a single Dirac cone. This can be seen, for example, from the cut at $k_y = -0.05$ Å$^{-1}$ in Fig. 4d, which reveals two partially overlapping cone-like dispersions. Indeed, the in-plane dispersion of these Dirac states (Fig. 4c) reveals rich structures related to these bulk states, comprising the central hyperbola around $k_y = 0$ Å$^{-1}$, which has a bow tie-like Fermi contour, and additional hyperbolic cones centered at around $k_y = \pm 0.15$ Å$^{-1}$ (see also Supplementary Fig. S7) which form asymmetric arcs. The spacing of this latter behavior is compatible with the periodicity imposed by the 6×1 phase, but their unusual distributions and the origin of the additional central bow tie structure remain unclear. Further detailed investigations including theoretical work will be required to clarify the nature of these states.

In summary, we have selectively stabilized the 6×1 charge ordered ground state of the layered topological semimetal IrTe$_2$ by employing uniaxial strain. The induced macroscopic domain sizes allow detailed insights into the electronic structure at the surface in both real and momentum space. Charge transfer in the strain-stabilized phase strongly reduces the out-of-plane Te bond strengths, electronically decoupling the layers and resulting in a Lifshitz transition, granting access to a previously inaccessible bulk Dirac dispersion that acts as the main inter-layer channel. Complementary measurements of the transport properties of the 6x1 phase, as well as of monolayer IrTe$_2$, are therefore highly desirable. We note that in contrast to the tensile strain utilized here, uniaxial compression may stabilize superconducting behavior[22–24] which, concomitant with the topological states, opens the possibility of strain-tunable topological superconductivity in IrTe$_2$.

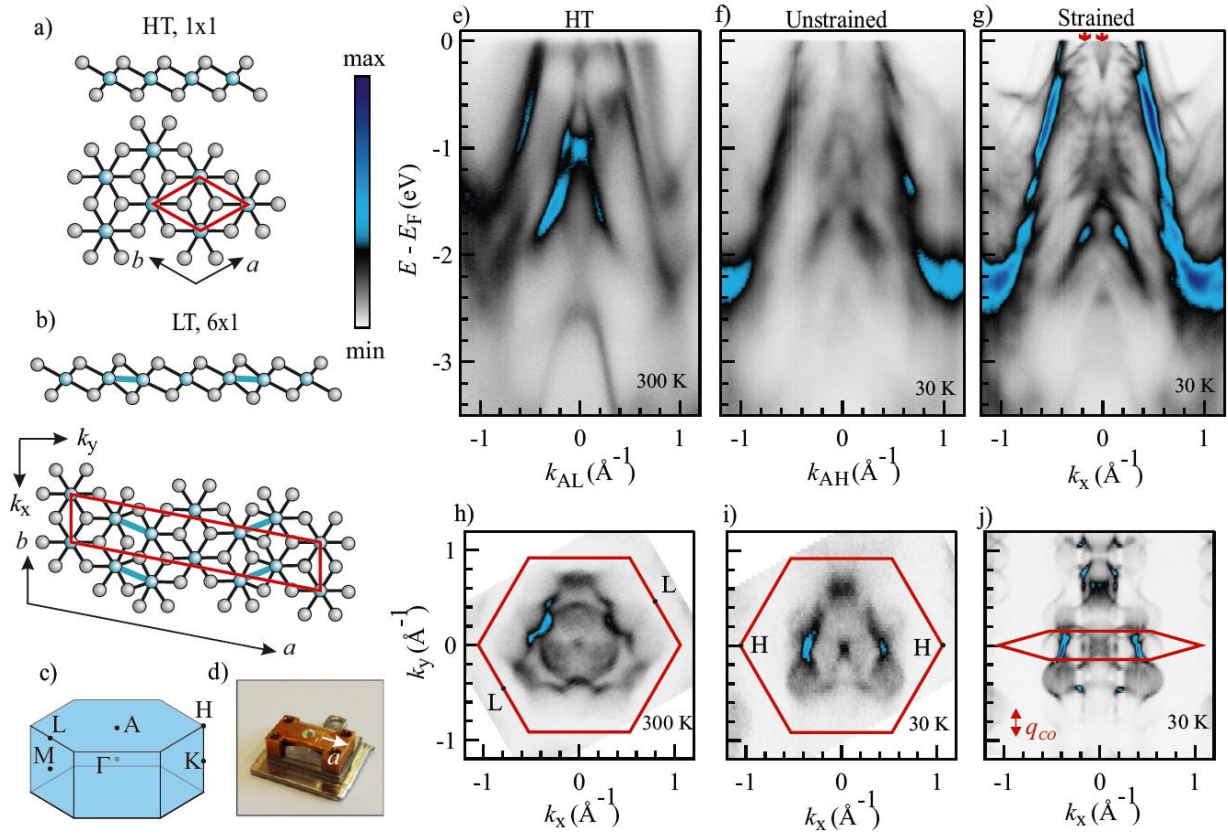

**Fig. 1. Effect of strain on the electronic structure of IrTe$_2$. a, b**, Top and side views of the crystal structure of IrTe$_2$ in the HT (**a**) and LT 6×1 phase (**b**). The formation of Ir-Ir dimers is highlighted by solid blue lines. Red lines are the unit cells of the corresponding phases. The experimental $k_x$ and $k_y$ directions are marked. **c**, Bulk Brillouin zone of the HT phase showing the high symmetry points. **d**, Photograph of the strain device. Samples are oriented to have strain along the *a*-direction of the HT phase. **e-g**, ARPES measurements ($h\nu = 90.5$ eV) of the HT unstrained sample obtained along the *LAL* direction of the bulk Brillouin zone (**e**), the LT ($T = 30$ K) unstrained sample obtained along *HAH* (**f**) and in the LT strained sample (**g**). The cut along $k_x$ in (**g**) corresponds to along the Ir-dimer chain direction i.e. the ×1 direction of the 6×1 unit cell (*b*-axis in real space). The intensity scale encodes the photoemission intensity measured in the experiment, which is proportional to the one-particle removal function, $A^-(k, \omega)$. **h-j**, corresponding Fermi surfaces for the three cases in (**e-g**). The phase transition breaks the three-fold rotational symmetry of the HT phase, but well-defined features appear only in the strained case. The corresponding Brillouin zones for the 1×1 and 6×1 phases are overlaid. Additional constant energy cuts are shown in Supplementary Fig. S1. Cuts along the $k_y$ direction (charge ordering direction, $q_{CO}$) revealing the 6×1 period are shown in Supplementary Fig. S2.

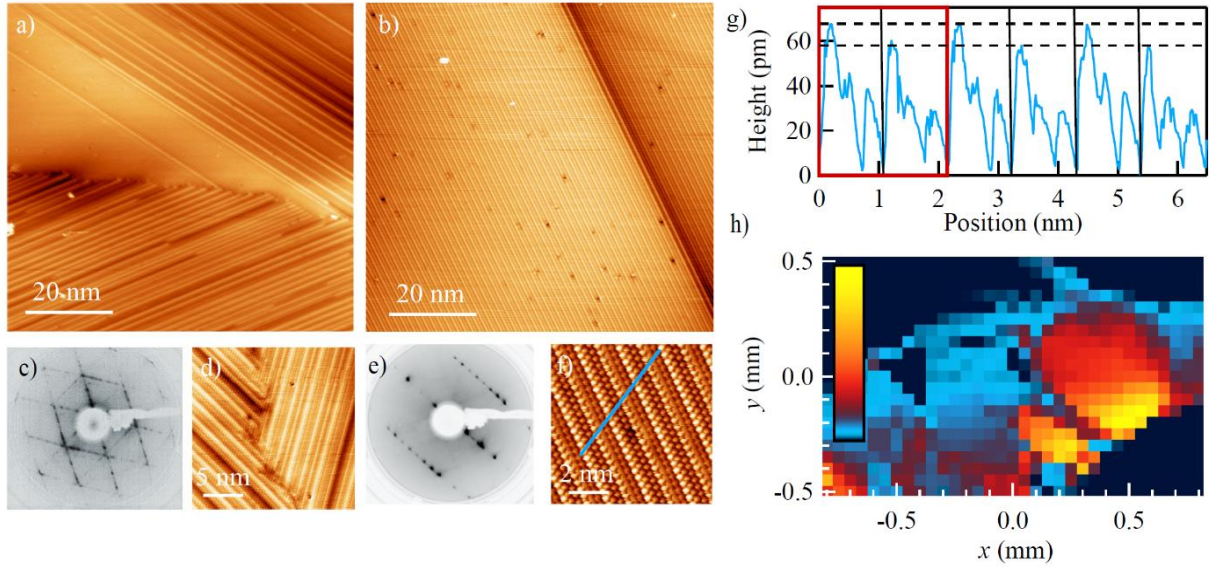

**Fig. 2. Real-space view of the strain-induced 6×1 phase. a**, **b**, Large scale STM images of an unstrained (**a**) and strained sample (**b**) at 4.5 K. A characterization of the strained sample across the sample surface is shown in Supplementary Fig. S3. **c**, **e**, LEED images of unstrained ($E = 85$ eV) (**c**) and strained samples ($E = 60$ eV) (**e**) at $T = 30$ K. A single rotational domain and 6×1 periodicity are observed in the LEED pattern of the strained sample. **d**, **f**, Atomic resolution STM images of the large scale images from **a** and **b**. The line cut marked in **f** is presented in **g** and reveals the 6× periodicity (red box) of the phase. Black vertical lines highlight inequivalent blocks of three atoms, which are differentiated by their respective heights (example horizontal dashed lines). **h,** Real-space ARPES intensity map ($h\nu = 6.3$ eV, $T = 30$ K) of a strained sample revealing strain-induced effects on the millimeter scale. The corresponding ARPES spectra obtained with a micro-focused laser is shown in Supplementary Fig. S4.

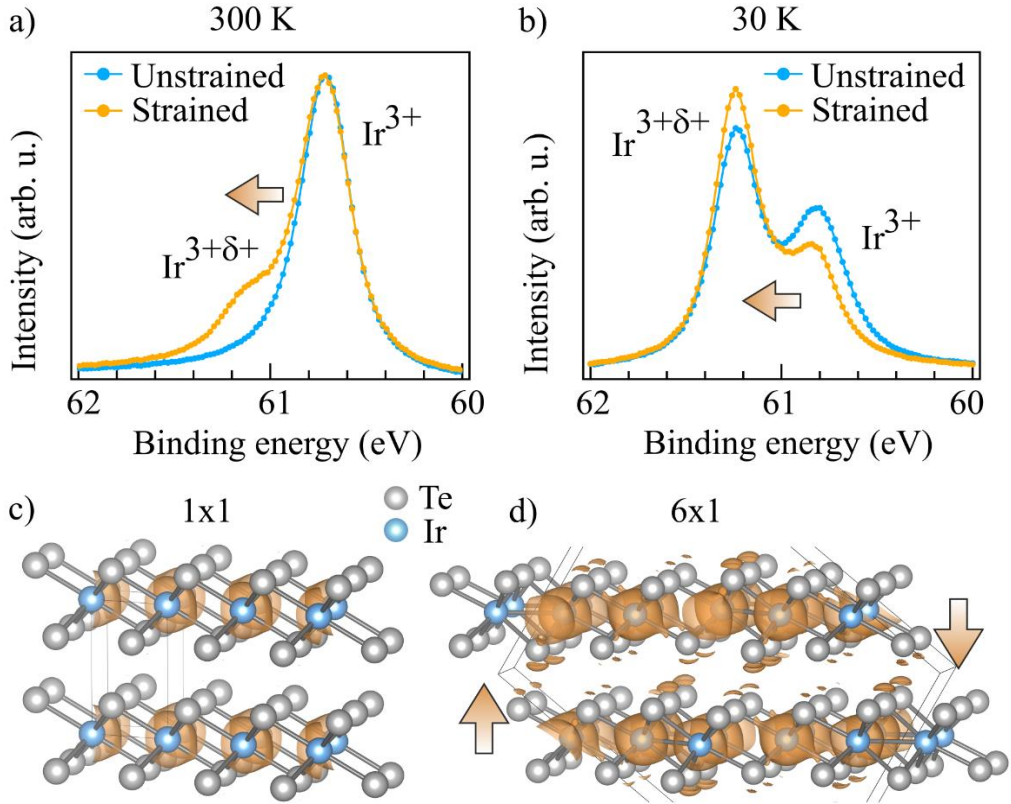

**Fig. 3. Strain induced charge transfer. a**, XPS measurements ($h\nu = 130$ eV, $T = 300$ K) of the Ir $4f_{7/2}$ core level for strained and unstrained samples, measured at the same position on strained and unstrained samples as the relevant ARPES data. The appearance of a distinct shoulder in the strained sample reveals the appearance of the Ir$^{3+\delta+}$ site, not normally present above 280 K, implying Ir to Te charge transfer. **b**, The same Ir $4f_{7/2}$ core level at 30 K. Again, the Ir$^{3+\delta+}$ peak is enhanced in the strained case. The ratio of the Ir$^{3+\delta+}$ peak area to the total weight, Ir$^{3+\delta+}$/(Ir$^{3+}$ + Ir$^{3+\delta+}$), is 0.67 in the 6×1 phase, as expected for 4 out of 6 atoms being dimerized, implying a pure 6×1 phase in the presence of strain. **c**, Calculated electronic density (orange isosurface) in the HT 1×1 phase. **d**, Electronic density in the 6×1 plotted at the same isosurface level as for the 1×1 phase in (**a**), revealing an increase of charge density in the inter-layer region (highlighted by arrows) that is maximized in the strain-stabilized 6×1 phase. Both calculations are obtained from the sum of electronic states between $E_F$ –7 eV and $E_F$.

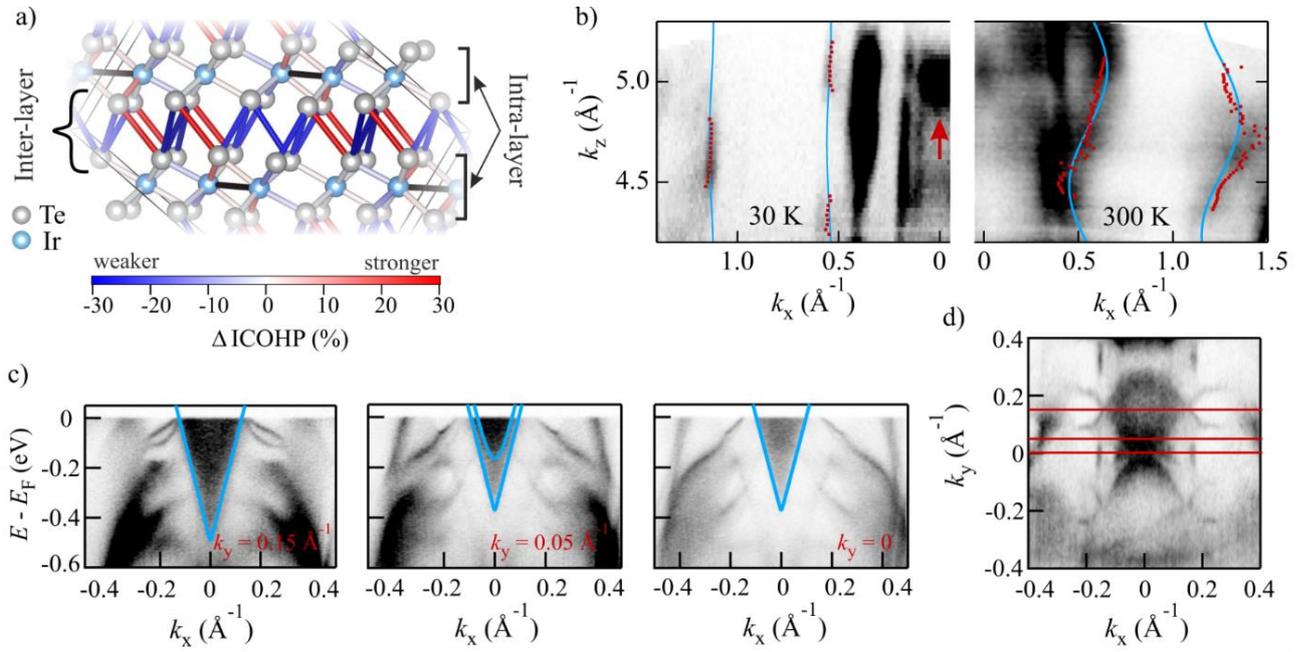

**Fig. 4. Observation of reduced inter-layer coupling and Lifshitz transition in the Dirac dispersions. a,** Calculated relative bond strengths in the 6×1 phase highlighting significant weakening of the inter-layer bonds. Bond strength changes are given with reference to the ICOHP in the HT phase and are caused by the charge transfer described in Fig. 3. The majority of inter-layer bonds are strongly weakened (blue) in the 6×1 phase. All bond strengths are tabulated in Supplementary Table S1. **b,** Out-of-plane ($k_z$) ARPES dispersions at $k_y = 0$ and $E_F$ for strained LT (left) and unstrained HT phases (right) (40 eV < $h\nu$ < 100 eV). The $k_x$ direction is along the Ir dimer chains, as in Fig. 1. Red markers show extracted contours obtained via fitting. Overlaid blue solid lines represent the resulting tight-binding dispersions. The observed narrowing of the warping along $k_z$ in the LT 6×1 phase implies a significant reduction of interlayer hopping, providing confirmation of the enhanced layer decoupling found in the calculation in **a**. The red arrow highlights the dispersive bulk-Dirac states (see also supplementary Fig. S6). **c,** ARPES spectrum ($h\nu = 20$ eV, $T = 30$ K) at $k_y = 0.15, 0.05$ and $0$ Å$^{-1}$ revealing overlapping bulk Dirac states. Overlaid solid blue lines are hyperbolic dispersions matched to the spectral weight. Additional spectra are presented in Supplementary Fig. 7. **d,** Fermi surface ($h\nu = 20$ eV, $T = 30$ K) highlighting the in-plane structure of the Dirac and surface states at the A-point. Red lines mark the positions of the dispersions presented in **c**.

**Methods**

**Sample growth and characterization.** Single crystals of IrTe$_2$ were grown using the self-flux method[32,46]. Samples were characterized by magnetic susceptibility and resistivity measurements[29], which confirmed the bulk phase transition temperatures of $T_{c1}$ = 278 K and $T_{c2}$ = 180 K in unstrained samples. Samples to be prepared for straining were chosen to have large flat areas with minimal cracks or flakes at the surface under an optical microscope to ensure as homogeneous as possible strain application. Bulk samples were initially cleaved with a scalpel to remove thicker layers, and then mounted onto the unstrained device and further thinned by Scotch tape cleaving.

**Strain device and characterization.** The strain device, shown in Fig. 1d, is a home-built design consisting of three parts: a molybdenum (Mo) base plate, a copper beryllium (CuBe) bridge, and a rounded aluminum (Al) block which is placed under the bridge. The maximum height of the Al block is machined to be slightly larger than the distance from the underside of the top of the CuBe-bridge to the surface of the base plate. Thus, when the pieces are screwed together, the CuBe-bridge if forced to bend by the Al block. Samples were oriented with Laue diffraction such that the bending axis was perpendicular to one of the three-fold symmetric directions in the HT phase i.e. along the *a*-axis of the HT phase (or equivalently, the *b*-axis). This is along the in-plane bond direction of half of the Ir-Ir dimers in the 6x1 unit cell. The sample was mounted on the CuBe-bridge of the strain device using a two-part epoxy (EPO-TEK E4110) which was cured and allowed to cool before strain was applied. Strain was applied manually by tightening the screws on the underside of the device, which connect the Mo base-plate with the Cu-bridge. To ensure maximally directed strain, the screws were tightened in pairs. All four screws were tightened loosely, following which the two screws on one side of the bridge were fully tightened. The remaining two screws were then gradually tightened in an alternating fashion, in order to allow as even an application of strain across the device as possible. The strain magnitude was calibrated using commercial strain gauges (Omega Engineering) with nominal resistance 350 Ω and gauge factor, $k$ = 2.2. The gauge was attached to an unstrained device of the design described above using the same epoxy as for the samples, and was then connected to a home-built Wheatstone bridge balanced circuit in a "quarter bridge" configuration. Together with a second (passive) gauge, this constituted one arm of the bridge. The second arm consisted of two 390 Ω resistors and a variable resistor (10 Ω) was used to balance the circuit. A source voltage of $V_s$ = 5 V was applied. Once balanced, the gauge was strained using the device and the output voltage was recorded with a Keithley digital multimeter. The typical output voltage induced by strain ($V_o$ = 4 mV) was well above the noise level (50 μV). The voltage output was converted to a strain value via[47]:

$$\varepsilon = \frac{4}{k}\frac{V_o}{V_s}$$

where ε, the total strain, is the sum of bending (tensile) strain and perpendicular (compressive) strain. In such a strain geometry, the perpendicular strain is considerably smaller than the bending strain[4], hence "strain" in the main text refers to the tensile bending strain. To separate further these components requires additional gauges to be placed on the underside of the device, which is impractical given the geometry and small size. From the above relation, we obtained the strain characteristics of the device. A maximum strain of up to 0.2 % was initially recorded during tightening due to plastic deformation of the CuBe-bridge. This relaxed to around 0.1 % once all screws were tight, which is therefore the maximum strain that could be applied to the sample using this particular device and Al block combination.

**Photoemission spectroscopy.** ARPES and XPS measurements were carried out at the I05 beamline[48] of Diamond light source, UK, with additional data obtained at the PEARL beamline[49] of the Swiss light source. Samples were cleaved in $10^{-9}$ mbar vacuum with Scotch tape at room temperature and then cooled using a liquid He flow cryostat at a rate of 5 K/min. Measurements were carried out in a base pressure of $10^{-11}$ mbar. The beam polarization used was linear horizontal (*p*-pol) and the beam size was 50 × 50 μm². A photon energy range of 20 - 100 eV was used for ARPES measurements, while XPS was carried out at 130 eV and 200 eV. Out-of-plane $k_z$ mapping was obtained by sweeping the incident photon energy through 40 eV < *hν* < 100 eV. The $k_z$ values were obtained using an inner potential of 12 eV. Spectra were acquired using a Scienta-Omicron R4000 photoelectron analyzer. Micro-ARPES mapping was carried out at the University of Fribourg. UV photons were generated using a commercial optical setup (Harmonix, APE GmbH) generating tunable output in the range 5.7 – 6.3 eV in non-linear crystals. Harmonic generation was driven by the output of a tunable OPO pumped by a 532 nm Paladin laser (Coherent, inc.) at 80 MHz. The sample surface was scanned by the encoded motion of a 6-axis cryogenic manipulator (SPECS GmbH). Spectra were acquired using a Scienta-Omicron DA30 analyzer.

**Scanning tunneling microscopy.** STM measurements were performed at the University of Fribourg on a commercial low-temperature STM (Scienta-Omicron) at 4.5 K in fixed current mode and with the bias voltage applied to the sample. Samples were cleaved in vacuum at $10^{-8}$ mbar pressure and measurements carried out in $10^{-11}$ mbar. Strain measurements utilized the same strain device design as for the ARPES measurements.

**Density functional theory.** The DFT calculations were performed using the projector augmented wave method[50,51] and the Perdew-Burke-Ernzerhof (PBE)[52] exchange-correlation functional within the VASP[53–56] code. The kinetic energy cutoff was set to 400 eV and a 5×15×4 k-point grid was used for Brillouin zone sampling. The starting structure for performing the structural relaxation in the 6x1 phase was the experimentally determined 6x1 structure[19] observed in $IrTe_{2-x}Se_x$ (space group C2/c, no. 15). The structure was relaxed until the forces were less than 1 meV/Å. The COHP[33] analysis was performed using the LOBSTER code[57–59]. Spin-orbit coupling was neglected in the calculations as the unit cell volumes with and without spin-orbit interaction (SOI) differ by only 0.8%. Similarly, differences in Ir-Ir distances are, at most, 1.2%. The density plots in Figs. 3 and 4 were generated with VESTA[60].

**Tight-binding model.** We describe the $k_z$ dispersion using the tight-binding model:

$$E_{\mathbf{k}} = -2t_a \cos(k_x a) - 2t_c \cos(k_z c) - \mu$$

where $t_a$ and $t_c$ are the energies associated with in-plane hopping along the Ir dimer chain direction and out-of-plane hopping, respectively; $a = 3.93$ Å and $c = 5.39$ Å are the lattice parameters along the corresponding directions; and $\mu$ is the chemical potential. Following a previously described procedure[34] we use the band energy minima at $k_x = 0$ ($E_\Gamma = -1.5$ eV), the Fermi wave vector at $k_z = 0$, equivalent to $k_\Gamma = 4.5$ Å$^{-1}$ ($k_{F,\Gamma} = 0.63$), and the Fermi surface warping extracted from Fig. 3b to determine the relations between the parameters and extract the hopping energies for the HT and LT phases. With $\mu = -0.45 - 2t_c$ and $t_a = -0.53$ eV, the extracted out-of-plane hopping value is $t_c = -0.156$ eV in the unstrained HT phase, and $t_a = -0.014$ eV in the LT strained phase.

**Acknowledgements.** This project was supported through the Swiss National Science Foundation (SNSF), Grant No. P00P2_170597. We gratefully acknowledge beam time from Diamond light source (proposal SI24880, beamline I05) and the Swiss light source (proposal 20170698, PEARL beamline). We thank P. Aebi for access to the photoemission and STM setups at the University of Fribourg and for helpful discussions during development of the strain device. Fruitful discussions with F. Baumberger during the initial phase of the project are warmly acknowledged. We acknowledge J. Chang for access to the Laue diffractometer at the University of Zurich, and J. Choi for technical support. We thank R. Ernstorfer for taking time to provide critical feedback on the manuscript. A.P. acknowledges the Osk. Huttunen Foundation for financial support, and the CSC-IT Center for Science, Finland, for computational resources. The work at the University of Zurich was supported by the Swiss National Science Foundation under Grant No. PZ00P2_174015. Skillful technical support was provided by O. Raetzo, B. Hediger and F. Bourqui.

**Author contributions.** ARPES and XPS measurements at Diamond light source were carried out by C.W.N., M.R., G.K., T.K., S.M., C.C. and C.M.. Laser ARPES measurements at the University of Fribourg were performed by C.W.N., M.R., T.J. and C.M.. STM measurements were performed by B.S., M.-L.M. and B.H.. M.M. provided support for XPS and ARPES measurements performed at the PEARL beam line that were carried out by C.W.N., M.R., G.K. and C.M during the initial phase of the project. ARPES data was analyzed by C.W.N. and XPS data by M.R.. Samples were grown and characterized by K.Y.M and F.O.vR. Charge density and bond strength DFT calculations were performed by A.P.. The project was initiated and managed by C.W.N. and C.M.. The manuscript was written by C.W.N. and C.M. with input and suggestions from all authors.

**Data availability:** Data are available from the authors upon reasonable request.

**Competing interests.** The authors declare no competing interests.

# Supporting materials for:

# Observation of a uniaxial strain-induced phase transition in the 2D topological semimetal IrTe$_2$

C. W. Nicholson[1,*], M. Rumo[1], A. Pulkkinen[1,2], G. Kremer[1], B. Salzmann[1], M.-L. Mottas[1], B. Hildebrand[1], T. Jaouen[1,3], T. K. Kim[4], S. Mukherjee[4], K. Y. Ma[5], M. Muntwiler[6], F. O. von Rohr[5], C. Cacho[4] and C. Monney[1,*]

[1]*Université de Fribourg and Fribourg Centre for Nanoscience, Chemin du Musée 3, CH-1700 Fribourg, Switzerland*

[2]*School of Engineering Science, LUT University, FI-53850, Lappeenranta, Finland*

[3]*Univ Rennes, CNRS, IPR (Institut de Physique de Rennes) - UMR 6251, F-35000 Rennes, France*

[4]*Diamond Light Source, Harwell Campus, Didcot, OX11 0DE, United Kingdom*

[5]*Department of Chemistry, University of Zurich, Winterthurerstrasse 190, CH-8057 Zurich, Switzerland*

[6]*Paul-Scherrer-Institute, Forschungsstrasse 111, CH-5232 Villigen PSI, Switzerland*

[*]Corresponding author: christopher.nicholson@unifr.ch (C.W.N); claude.monney@unifr.ch (C.M.)

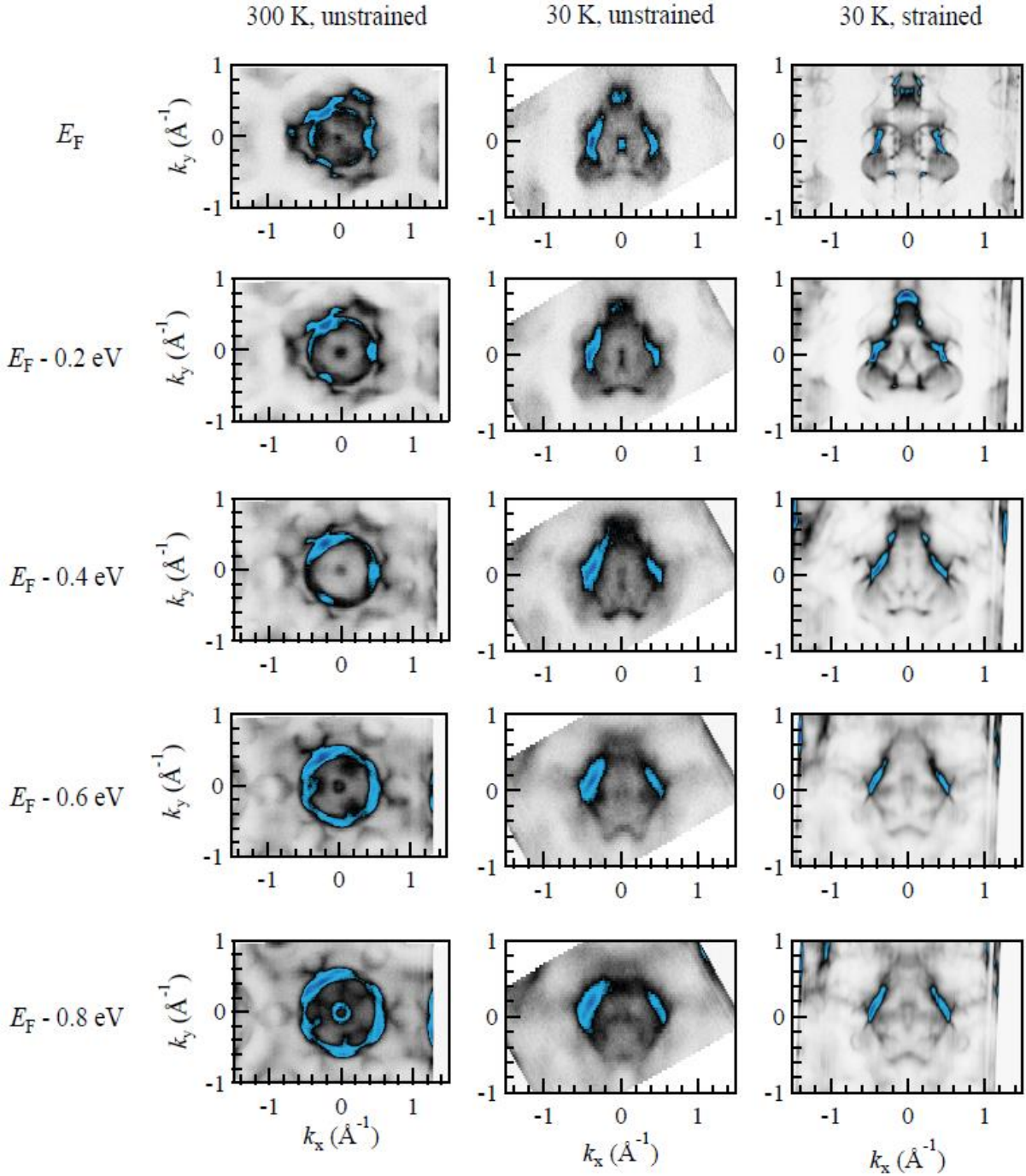

Fig. S1. **Constant energy maps of unstrained and strained IrTe$_2$ samples.** Evolution of the electronic structure in the ($k_x$, $k_y$)-plane over an energy range of 0.8 eV below $E_F$ ($h\nu$ = 90.5 eV). **Left column**, unstrained sample in the HT phase. **Middle column**, unstrained sample in the LT phase showing a breaking of the rotational symmetry but very diffuse features as a result of a lack of a single well-defined periodicity. **Right column**, strained sample in the LT phase revealing sharp features as a result of the strain stabilized single phase. The directionality and mirror plane visible at the Fermi level in the strained sample are maintained also at lower binding energies.

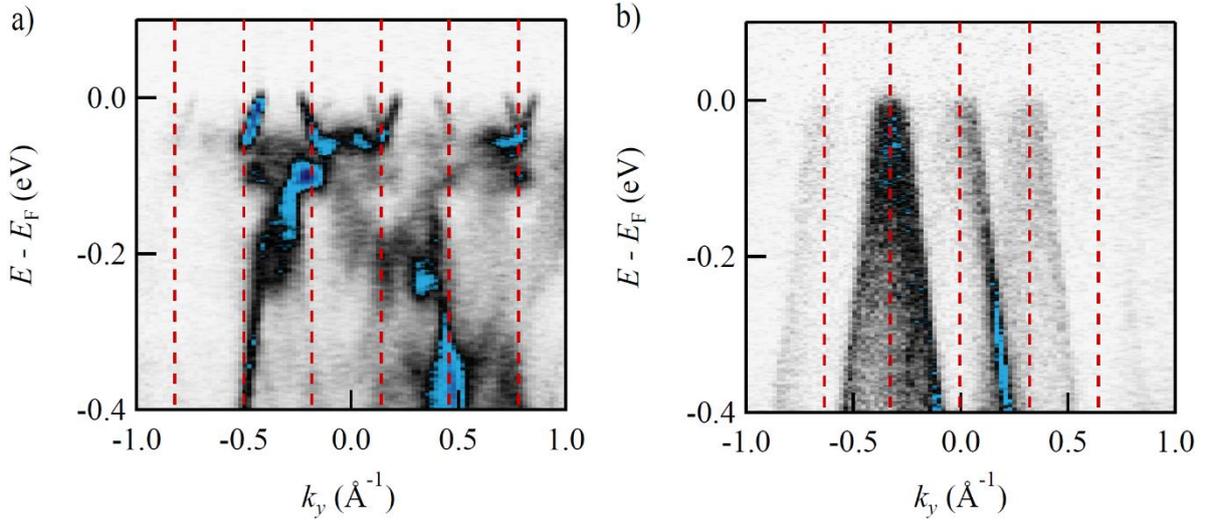

Fig. S2. **ARPES evidence for the 6x1 periodicity in IrTe$_2$. a**, cut through the Fermi surface presented in the main text Fig. 1j along $k_y$ at $k_x = 0.2$ Å$^{-1}$ revealing repeated parabolic surface states dispersing below $E_F$. Red dashed lines mark a distance of 0.33 Å$^{-1}$, which within our experimental sensitivity corresponds to the 6x periodicity of the LT phase (0.35 Å$^{-1}$). **b**, cut as in **a** obtained at $k_x = -0.6$ Å$^{-1}$ passing through the bulk states revealing the same periodicity as in (**a**).

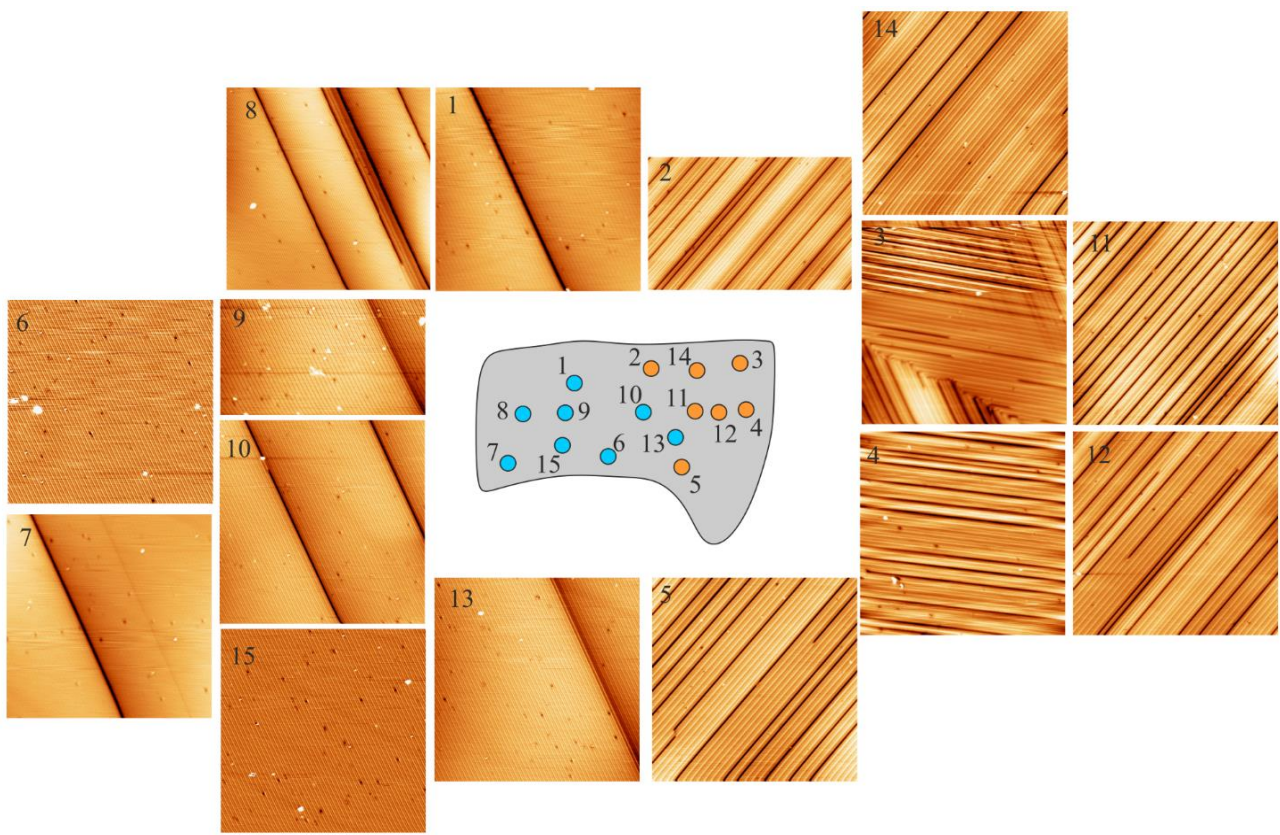

Fig. S3. **STM characterisation of the strained IrTe₂ sample surface.** STM images obtained at multiple positions across a strained sample surface, as marked numerically on the schematic sample area in the centre. The large single domains of the 6x1 phase are evident across more than half of the sample. The right side of the sample shows a mixture of orientations and periodicities. Images are 100 x 100 nm$^2$, except for 2 and 9 which are 100 x 60 nm$^2$.

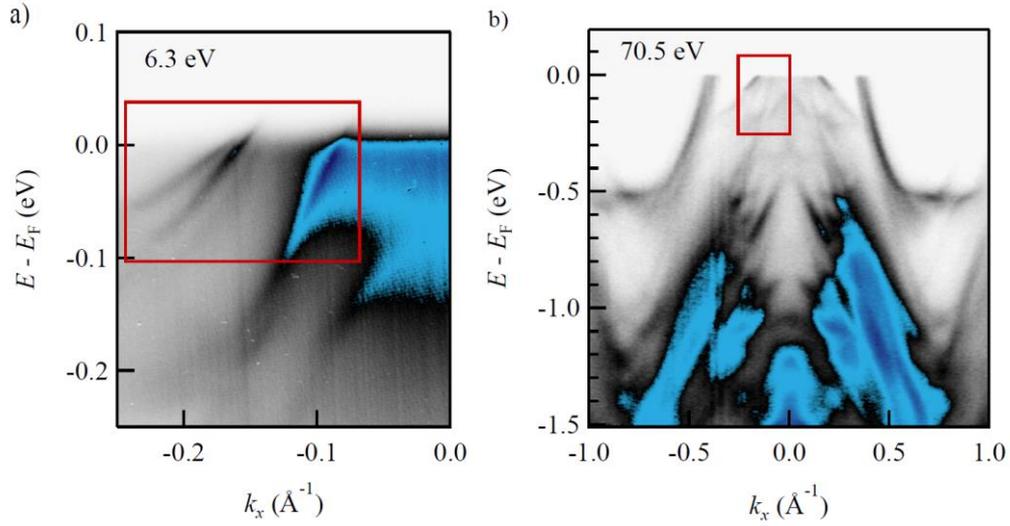

Fig. S4. **Spectra of strained IrTe$_2$ samples used for ARPES mapping. a**, Laser ARPES spectra obtained close to the Γ-point of the bulk Brillouin zone ($h\nu$ = 6.3 eV). The red rectangle highlights the region summed over in order to obtain the mapping in Fig. 2h of the main text. **b**, The ARPES spectra obtained directly at the Γ-point of the bulk Brillouin zone ($h\nu$ =70.5 eV). The red rectangle represents the area of the data obtained with the laser in (**a**). In contrast to the spectra obtained at the bulk A-point (main text Fig. 1) the bulk Dirac states are not observed.

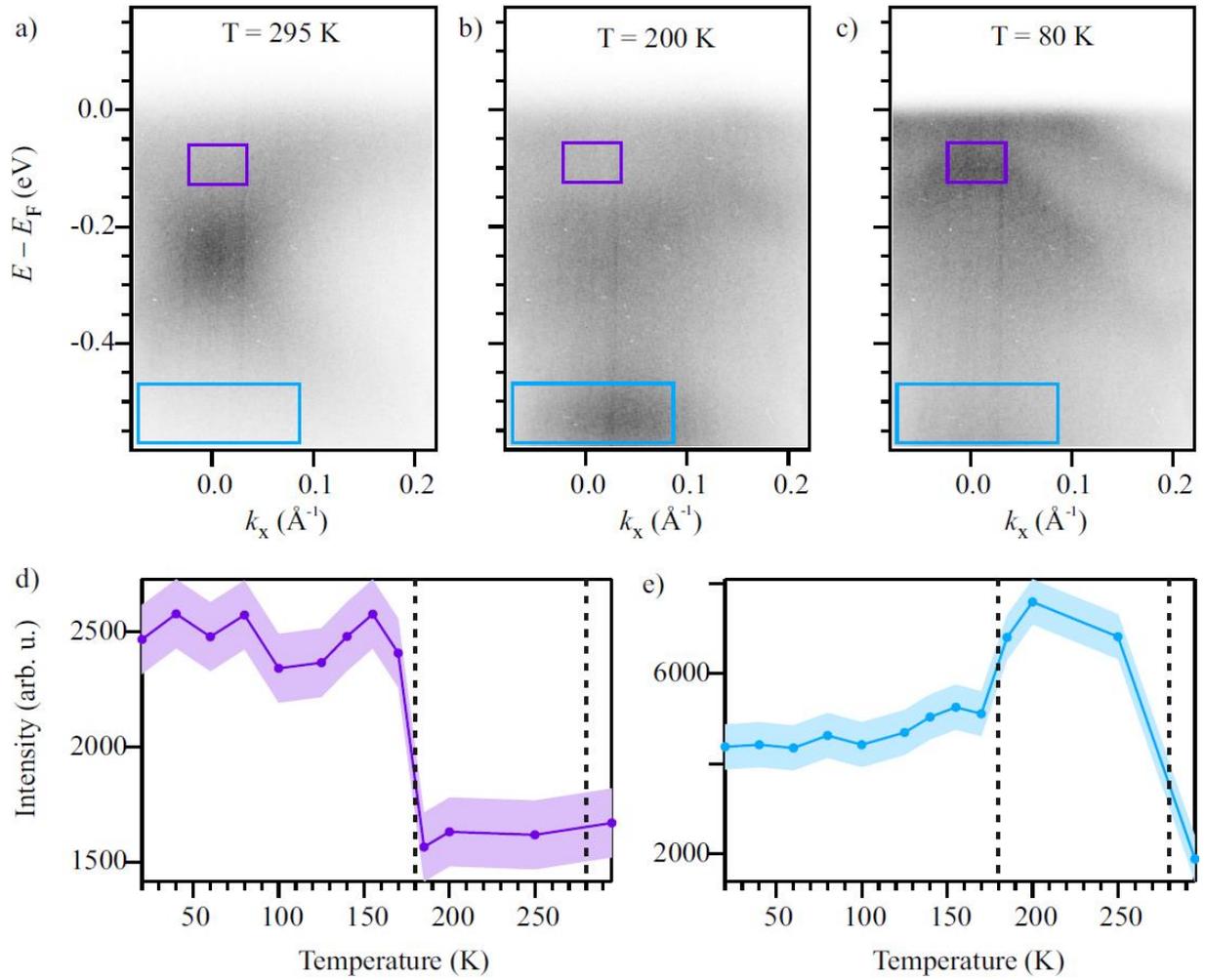

Fig S5. **Temperature dependence of the electronic structure in a strained sample.** Representative ARPES spectra obtained close to the Γ-point of the bulk Brillouin zone ($h\nu$ = 6.3 eV) at (**a**) 295 K in the 1x1 phase, (**b**) at 200 K in the 5x1 phase and (**c**) at 80 K in the 6x1 phase. In order to ensure that we probe the evolution of a strained region, we have first cooled the sample to 30 K and searched for the clear 6x1 features as seen in Fig. S4a. The sample was then gradually warmed into the 1x1 phase following which temperature dependent measurements were made during cooling on the strained region. This temperature cycle reduces the surface quality, thus we focus on bulk bands to characterize the temperature dependent behavior. We clearly observe an abrupt transition to the 6x1 phase at ~180 K, ruling out a continuous transition, or a significant change to the transition temperature at this level of strain. The density of data points at higher temperatures does not allow a definitive statement on whether or not an abrupt transition occurs at 280 K, but we have observed no indications of an altered transition to the 5x1 phase.

Te-Te (1x1)

| d (Å) | ICOHP (eV) |
|---|---|
| 3.65571 | -0.25326 |

Te-Te (6x1)

| d (Å) | ICOHP (eV) | Δ ICOHP (%) |
|---|---|---|
| 3.5739 | -0.29835 | +17 |
| 3.68827 | -0.19679 | -22.2 |
| 3.72114 | -0.18435 | -27.2 |
| 3.77539 | -0.1782 | -29.6 |

Ir-Ir (1x1)

| d (Å) | ICOHP (eV) |
|---|---|
| 3.99042 | -0.07234 |

Ir-Ir (1x1)

| d (Å) | ICOHP (eV) | Δ ICOHP (%) |
|---|---|---|
| 3.11685 | -0.41966 | +580 |
| 3.9628 | -0.06564 | -9.3 |
| 4.0403 | -0.06923 | -4.2 |
| 4.0403 | -0.08841 | +22.3 |
| 4.08244 | -0.08649 | +19.5 |
| 4.12271 | -0.06502 | -10.1 |

Ir-Te (1x1)

| d (Å) | ICOHP (eV) |
|---|---|
| 2.68013 | -2.09089 |

Ir-Te (6x1)

| d (Å) | ICOHP (eV) | Δ ICOHP (%) |
|---|---|---|
| 2.62856 | -2.27526 | +8.8 |
| 2.65961 | -2.1296 | +1.8 |
| 2.66961 | -2.10104 | +0.48 |
| 2.67566 | -2.09049 | -0.02 |
| 2.67641 | -2.07541 | -0.7 |
| 2.68189 | -2.05422 | -1.8 |
| 2.70557 | -1.95366 | -6.5 |
| 2.71578 | -1.97177 | -5.6 |
| 2.75275 | -1.80736 | -13.5 |

Table S1. **Results of bond calculations for IrTe$_2$.** Calculated inter-atom distances ($d$) and integrated crystal orbital Hamiltonian potential (ICOHP) (bond strength) values for the three possible bond types in the HT 1x1 and LT 6x1 phases. Larger negative values correspond to stronger bonds. The change of bond strength relative to the HT phase is expressed as a percentage in the third column (Δ). A positive increase corresponds to an increase in bond strength and related to the color scale in Fig. 3d of the main text. The strongest bonds are clearly between in-plane Ir-Te, while with the exception of the dimerized Ir-Ir bond, all other Ir-Ir bonds are so weak as to be essentially non-bonding. The most significant changes occur in the out-of-plane Te-Te bonds, thus affecting the dimensionality of the related states.

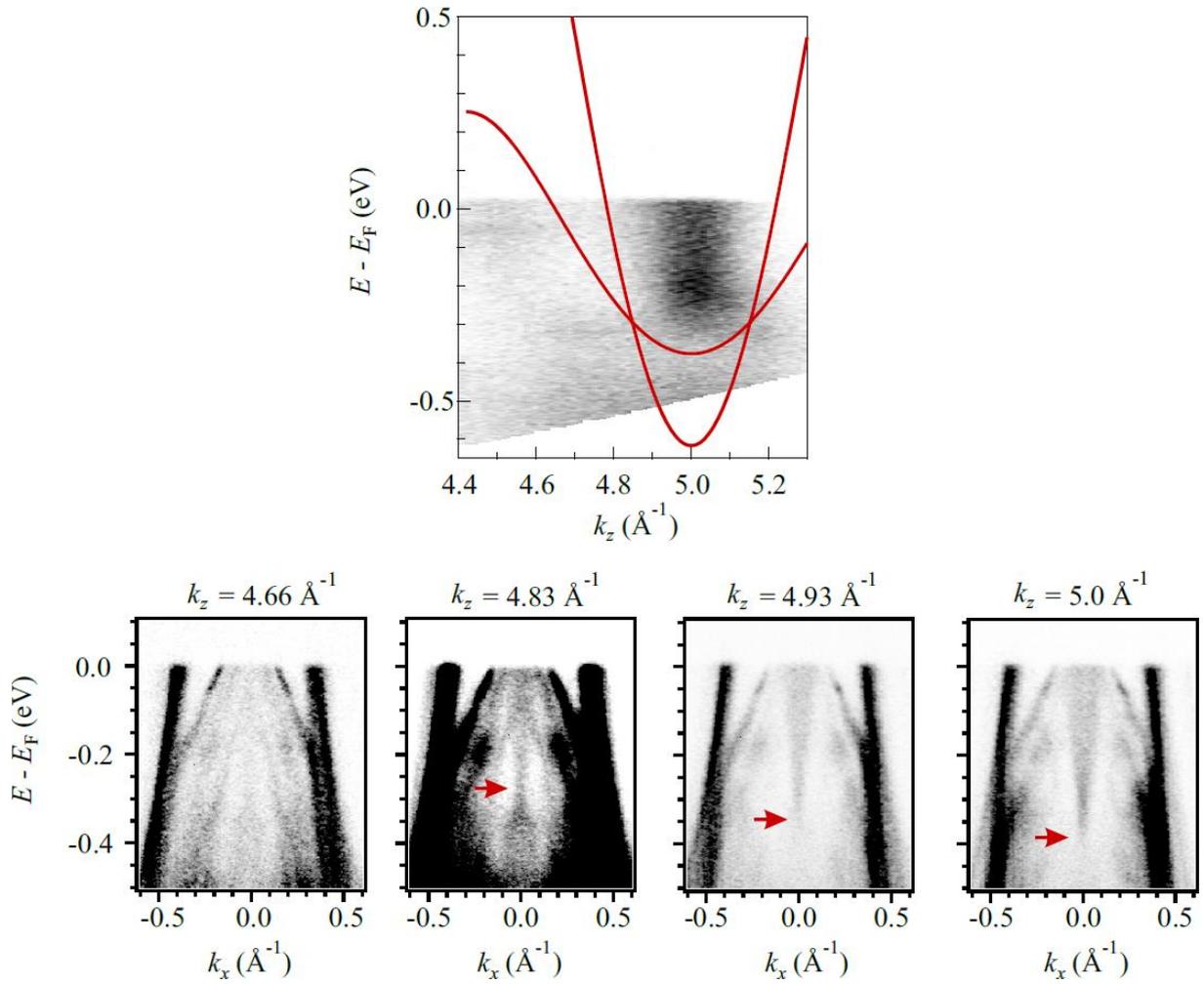

Fig. S6. **Out-of-plane dispersion of the bulk Dirac states in the 6x1 phase. Upper panel**, ARPES spectra ($T = 30$ K) in the $k_z$ (inter-layer, Γ-A) direction at $k_x = k_y = 0$, obtained by sweeping the incident photon energy through 65 eV $< h\nu <$ 100 eV. The Γ (A) point are determined to be close to 4.5 Å$^{-1}$ (5.0 Å$^{-1}$). The overlaid curves are the dispersions along Γ-A calculated in the 1x1 phase, with the chemical potential set to have the minimum of the upper band at 5 Å$^{-1}$ at the minima of the data. States close to 5 Å$^{-1}$ correspond to the bulk Dirac states discussed in the main text and are observed to disperse strongly, implying 3D character. **Lower panel**, dispersions along the Ir dimer chain direction ($k_x$) at selected $k_z$ positions. The out-of-plane dispersion of the bulk Dirac states is evident from the shift of the band minima, marked by arrows. Images have been saturated to highlight the Dirac dispersions. The bulk Dirac point is observed close to 4.83 Å$^{-1}$.

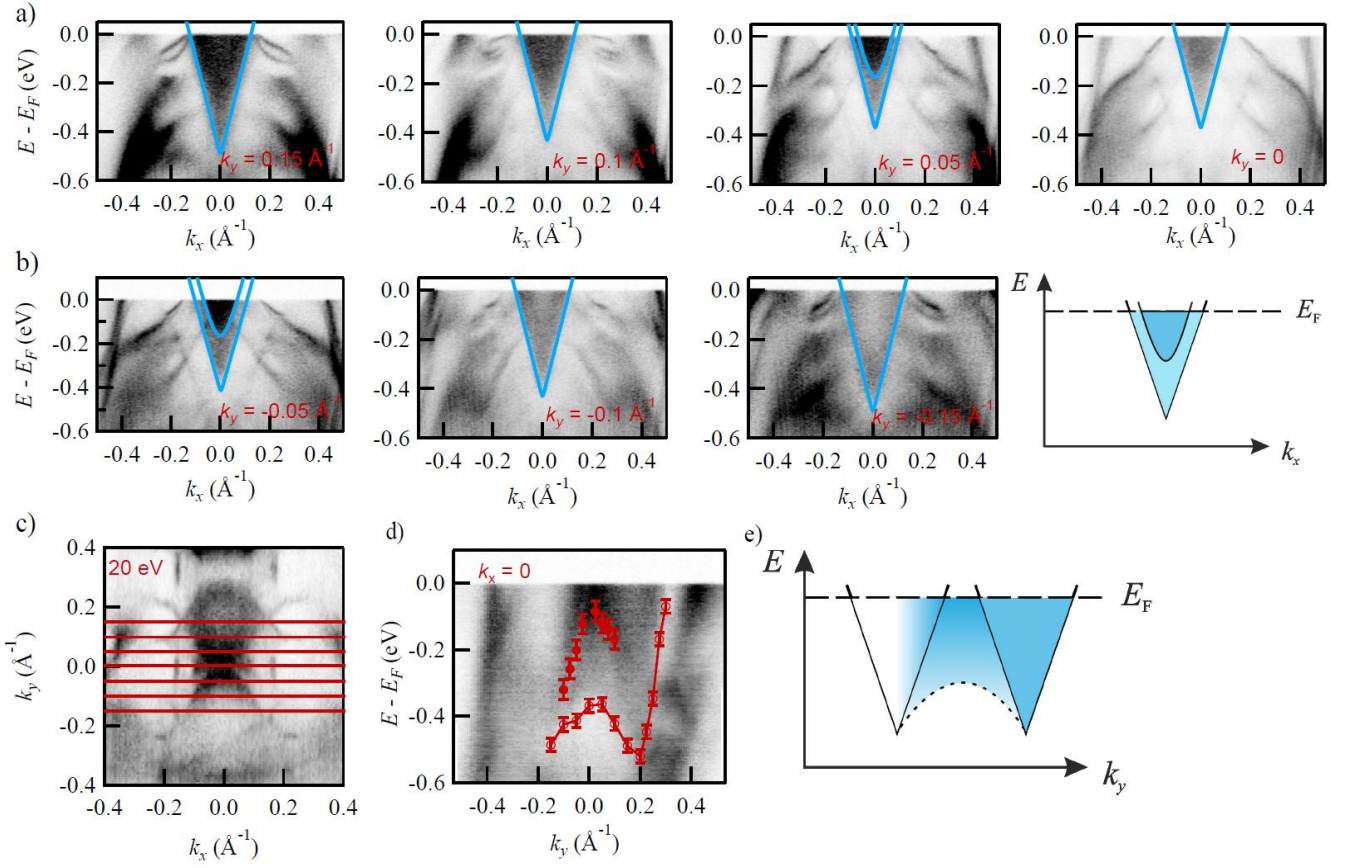

Fig. S7. **In-plane dispersion of the bulk Dirac cone in IrTe$_2$. a**, ARPES cuts along $k_x$ at different $k_y$ values as marked in the relevant panel from 0.15 Å$^{-1}$ to 0 (at the bulk Brillouin zone A-point, $h\nu$ = 20 eV). Hyperbolic dispersions for conical dispersions are overlaid to highlight the Dirac-like nature of the states. **b**, ARPES cuts as in a for the $k_y$ range -0.05 Å$^{-1}$ to -0.15 Å$^{-1}$ ($h\nu$ = 20 eV). The double hyperbolic dispersion shown schematically in the last panel can be most clearly observed at $k_y$ = ±0.05 Å$^{-1}$ as highlighted by the two overlaid dispersions in the respective panels. In total, three Dirac-like dispersions are observed as described in the main text: the central dispersion at $k_y$ =0 and one on either side. **c**, Fermi surface of the strained LT phase at the A-point of the bulk Brillouin zone ($h\nu$ = 20 eV) showing the position of the cuts in (**a**) and (**b**). **d**, ARPES spectra and minima positions of the hyperbolic dispersions as a function of $k_y$ as extracted from the panels in (**a**) and (**b**). The cones at $k_y$ = ±0.15 Å$^{-1}$ are compatible with the periodicity imposed by the 6x1 phase, but does not explain the continued presence of the states at ($k_x$, $k_y$) = 0. Curiously, the $k_y$ behavior of the central dispersion is itself not cone-like as can be seen from its curved minima position (**d**). **e**, Schematic of the unusual concave band minima of the central dispersion in between the two flanking cones.